\documentclass[pra,twocolumn,superscriptaddress,amsmath,amssymb]{revtex4-2}
\usepackage{amsmath}
\usepackage{amsfonts}
\usepackage{amssymb}
\usepackage{graphicx}
\usepackage{color}
\usepackage{txfonts}
\usepackage{float}
\usepackage[titletoc]{appendix}
\usepackage[colorlinks={true}]{hyperref}
\hypersetup{citecolor={blue}, filecolor={blue}, linkcolor={blue}, urlcolor={blue}}
\begin{document}


\title{\textbf{Precise Quantum Control of Molecular Rotation Toward a Desired Orientation} 
}%

\author{Qian-Qian Hong}
\affiliation{
 Hunan Key Laboratory of Nanophotonics and Devices, Hunan Key Laboratory of Super-Microstructure and Ultrafast Process, School of Physics, Central South University, Changsha, 410083, China
}
\author{Daoyi Dong}
\affiliation{Australian Artificial Intelligence Institute, Faculty of Engineering and Information Technology, University of Technology Sydney, NSW 2007, Australia}
\author{Niels E. Henriksen}
\affiliation{
Department of Chemistry, Technical University of Denmark, Building 207, DK-2800 Kongens Lyngby, Denmark
}
\author{Franco Nori}
\affiliation{
Quantum Computing Center and Theoretical Quantum Physics Laboratory, RIKEN, Saitama 351-0198, Japan
}
\affiliation{
Physics Department, University of Michigan, Ann Arbor, Michigan 48109, USA
}
\author{Jun He}
\email{junhe@csu.edu.cn}
\affiliation{
Hunan Key Laboratory of Nanophotonics and Devices, Hunan Key Laboratory of Super-Microstructure and Ultrafast Process, School of Physics, Central South University, Changsha, 410083, China
}
\author{Chuan-Cun Shu}
\email{cc.shu@csu.edu.cn}
\affiliation{
Hunan Key Laboratory of Nanophotonics and Devices, Hunan Key Laboratory of Super-Microstructure and Ultrafast Process, School of Physics, Central South University,
Changsha, 410083, China
}


\date{\today}

\begin{abstract}
The lack of a direct map between control fields and desired control objectives poses a significant challenge in applying quantum control theory to quantum technologies. Here, we propose an analytical framework to precisely control a limited set of quantum states and construct desired coherent superpositions using a well-designed laser pulse sequence with optimal amplitudes, phases, and delays.  This theoretical framework that corresponds to a multi-level pulse-area theorem establishes a straightforward mapping between the control parameters of the pulse sequence and the amplitudes and phases of rotational states within a specific subspace. As an example, we utilize this approach to  generate 15 distinct  and  desired  rotational superpositions of  ultracold polar molecules, leading to 15 desired field-free molecular orientations. By optimizing the superposition of the lowest 16 rotational states, we demonstrate that this approach can achieve a  maximum orientation value of $|\langle\cos\theta\rangle|_{\rm{max}}$ above 0.99, which is very close to the global optimal value of 1 that could be achieved in an infinite-dimensional state space. This work marks a significant advancement in achieving precise control over multi-level subsystems within molecules. It holds potential applications in molecular alignment and orientation, as well as in various interdisciplinary fields related to the precise quantum control of ultracold polar molecules, opening up considerable opportunities  in molecular-based quantum techniques.  
\end{abstract}

\maketitle


\emph{Introduction.---} Owing to the inclusion of vibrational and rotational degrees of freedom, molecular systems offer exceptional opportunities for encoding quantum information beyond the capabilities of other platforms   \cite{2002PRL_DeMille,2012PRL_Manai,2022PRX_Wang,2023Sci_Yi}. Advancements in experimental techniques, theoretical models, and computational methods have led to significant progress in quantum control of molecular rotation in recent years \cite{2019NC_Chatterley,2019RMD_Sugny,2021PRR_Zak,2022NC_Mullins,2022PRL_Wu,2024PRA_Milner}. One of the main goals in this field is to develop advanced methods for achieving molecular orientation without the need for external static fields \cite{2001PRL_Niels,2010PRL_Sakai,2011PRL_Fleischer,2014PRL_Jones,2018NC_Wu,2020PRL_Xu,2023PRL_Fan}. The physical mechanism for this phenomenon involves generating rotational wave packets consisting of both even and odd angular momentum quantum numbers. This capacity  holds immense potential across a diverse range of applications, including photochemistry  \cite{2017NP_Moses,2023PRA_Fukahori}, precision spectroscopy  \cite{2010PRL_Leroux,2018NP_Alighanbari}, ultrafast science  \cite{2011PRA_Rubio, 2019PRA_Li,2021PRR_Zhou,2022PRL_Rezvan}, molecular imaging  \cite{2018PRL_James,2021PRApplied_Qi, 2019OE_Agarwal}, quantum information processing  \cite{2005JMO_Shapiro}, and material science  \cite{2012NM_Wu,2023PRL_Karle}.\\ \indent  Quantum control techniques, particularly with the use of shaped laser pulses, have made it possible to harness the rotational states of molecules as the fundamental units, referred to as qubits, for performing quantum computing and simulations  \cite{2017PR_Nori,2020PRX_Albert,2009CP_Yamashita,2014RMP_Nori,2020PRA_Michael,2024PRA_Nori}. This approach allows for the encoding, manipulating, and measuring of qubits within a finite-dimensional space of rotational states  \cite{2011JCP_Wei,2018QST_Blackmore,2018CS_Ni,2019NJP_Phelan}. In light of this progress, an intriguing question arises: Is it feasible to exert precise control over a specific subset of rotational states to achieve a desired field-free orientation of molecules? It requires accurately determining the amplitude and phase of the wave function for each involved rotational state. Despite significant progress in experimental techniques and theoretical methods for controlling quantum systems \cite{2011PRA_Ren-Bao,2015EPJD_Sugny,2017N_Chou,2022SA_Kallush,2023_Dong}, establishing a straightforward analytical approach to uncover the intricate relationship between control fields and rotational wave functions remains a complex task in quantum control. This complexity can be attributed to the involvement of multiple rotational states and complex transitions.\\ \indent 
In this Letter, our primary goal is to establish an analytical framework that can effectively address the complexities of achieving precise control over a given number of rotational states and ultimately achieve any desired rotational superposition. Through the analytical design of a pulse sequence, we derive a general multi-level pulse-area theorem that accurately maps characteristic parameters of each subpulse, such as amplitude, phase, time delay, and center frequency, onto the excited rotational wavepacket of molecules.  To validate our theoretical method,  we perform numerical simulations to confirm that the analytically designed pulse sequence can achieve precise control over a finite number of rotational states of ultracold dipolar molecules in their absolute ground state and work towards attaining a desired field-free molecular orientation. This theoretical framework holds promising implications for molecular-based quantum techniques.\\ \indent
\begin{figure*}[ht]
\centering
\resizebox{0.85\textwidth}{!}{%
\includegraphics{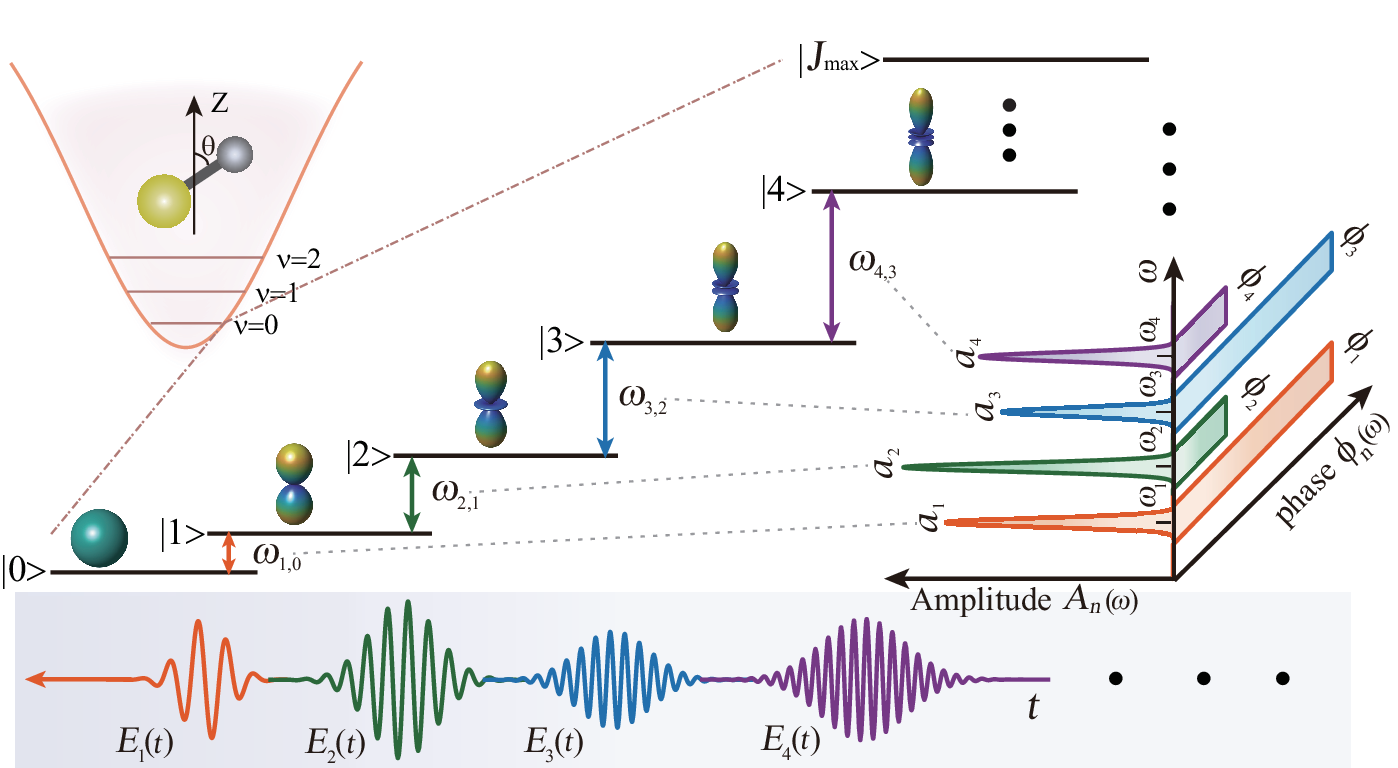} }\caption{Schematic illustration of generating a desired coherent superposition of rotational states blueof ultracold polar molecules initially in the absolute
ground state ($\nu=0$ and $J=0$) of the ground electronic potential by a precisely designed pulse sequence. The subspace consists of a finite number of rotational states, which is driven from the ground state $|0\rangle$ to the highest rotational state $|J_{\rm{max}}\rangle$ step by step by a time-domain optimized  pulse sequence, analytically designed in the frequency domain by control parameters $\left\{a_n, \phi_n, \omega_n \right\}$.} \label{fig1}
\end{figure*}
\emph{Precise Control Method.---} We now explain how to establish a general precise control method based on pulse sequence excitation. 
Figure \ref{fig1} illustrates our control scheme for manipulating rotational quantum states of ultracold dipolar molecules using a linearly polarized pulse sequence
\begin{equation} \label{pulse}
E(t)\equiv\sum_{n=1}^{N}E_n(t)=\sum_{n=1}^{N}\mathcal{E}_nf_n(t)\cos\left [ \omega_n(t-\tau_{n})+\phi_{n} \right ],
\end{equation}
where $\mathcal{E}_n$ denotes  the electric field's amplitude, $f_n(t)$ the envelope function, $\omega_n$ the center frequency, $\tau_{n}$ the center time, $\phi_{n}$ the phase of the $n$th subpulse $E_n(t)$, and $N$ the total number of subpulses. By carefully designing each subpulse with specific parameters, the objective of this approach is to transfer the molecules  from the  ground rotational state $|J=0\rangle$ in the ground vibrational state of the ground electronic state at the initial time $t_0$ to a target  subspace of rotational states at the final time $t_f$ defined by ($\hbar=1$)
\begin{equation}\label{target}
    \left| \psi(t_f) \right \rangle = \sum_{J=0}^{J_{\mathrm{max}}} c_J \exp(i\varphi_J)\exp(-i\omega_Jt_f)\left | J \right \rangle,
\end{equation}
where $\omega_J$ represents the eigenfrequency, $c_{J}$ the real positive amplitude, $\varphi_J$ the phase of the rotational state $|J\rangle$ with rotational quantum number $J$, and $J_{\mathrm{max}}$ the highest rotational state in the subspace.\\ \indent 
To establish a mapping between the control parameters $\{\mathcal{E}_n, \omega_n, \tau_{n}, \phi_n\}$ and the quantum state parameters $\{c_J, \varphi_J\}$, we elucidate our approach by investigating the impact of a single pulse on a two-level system, commonly known as a single qubit. This analysis led to the identification of Rabi oscillations and the concept of the $\pi$-pulse \cite{2014PRL_Dominici,2017QST_Fischer,2021PRA_Shi,2023PR_Nori}. Here,  the two-level system is composed of $|0\rangle$ and $|1\rangle$ with a transition dipole moment  $\mu_{1,0}$ and a transition frequency  $\omega_{1,0} = \omega_1 - \omega_0$. By employing the first-order Magnus expansion of the unitary time-evolution operator, the time-dependent wave function of the system in the interaction picture reads  \cite{2022PRA_Guo}
\begin{equation}
    \left | \psi(t)\right\rangle_1 = \cos{\theta_1(t)}\left | 0\right \rangle + i\sin{\theta_1(t)}\exp\left[-i(\phi_1-\omega_1\tau_1)\right]\left | 1 \right \rangle,
\end{equation}
where the pulse area $\theta_1(t)$ is given by $\theta_1(t)=\mu_{1,0}\left|\int_{t_{0}}^{t}E_1(t')\exp(-i\omega_{1,0}t')dt'\right|$. By employing the frequency-domain analysis of the control pulse $E_n(t)=\frac{1}{\pi}\mathrm{Re}\Bigg[\int_{0}^{\infty}A_n(\omega)\exp[i(\phi_n-\omega\tau_n)]\exp(i\omega t)d\omega\Bigg]$, where $A_n(\omega)$ denotes the spectral amplitude and $\phi_n(\omega)$ represents the spectral phase, we can deduce that the pulse area $\theta_1(t_{f})$ is solely determined by $A_1(\omega)$ at $\omega_{1,0}$. By considering resonant excitation with $\omega_1=\omega_{1,0}$, we can ensure that the value of $\theta_1(t_f)$ is precisely controlled by the control parameter $\mathcal{E}_1$. As a result, this has given rise to a well-established two-level pulse-area theorem, which allows for precise manipulation of the control parameters $\mathcal{E}_1$ and $\phi_1$. This manipulation can be used to generate an arbitrary coherent superposition of two states, serving as the fundamental unit of quantum information and holding considerable significance in quantum techniques.\\ \indent 
While various quantum optimal control theory methods have been successfully applied to finding optimal external fields \cite{1990JMS,1998PRA_Ortigoso,2005JCP_Salomon,2016PRA_Shu,2019PRA_Shu}, deriving the analytical solution for the time-dependent wave function of a multi-level system is generally more complex than that of a two-level system \cite{1999PRA_Montesinos,2009OL_Groves}. This complexity necessitates the advancement of a multi-level pulse-area theorem for precise control of high-dimensional  quantum systems. We have made strides toward this  ultimate goal by examining various three-level models \cite{2019PRL_Shu,2021PRA_Hong,2023JPA_Fan}. Through the utilization of pure rotational ladder-climbing excitations between states, we established the mapping of control amplitudes $\mathcal{E}_1$ and $\mathcal{E}_2$, as well as control phases $\phi_1$ and $\phi_2$, onto the amplitudes and phases of three rotational states \cite{2021PRA_Hong}. Here, we analyze a pulse sequence that comprises a series of sub-pulses distributed at specific time intervals. By considering each subpulse that couples two adjacent rotational states independently, as depicted in Fig. \ref{fig1}, we derive a general solution of the time-dependent wave function comprising multiple rotational states  without invoking the rotating wave approximation (RWA)  \cite{SM} 
\begin{equation}\label{psi}
\begin{aligned}
    \left|\psi(t)\right\rangle_{N}=&\cos{\theta_1(t)  }\left | 0 \right \rangle+\sum_{J=1}^{N-1}\cos{\theta_{J+1}(t)}\prod_{n=1}^{J}i\sin{\theta_{n}(t) }\\ 
    &\times\exp\left[-i(\phi_n-\omega_{n,n-1}\tau_{n})\right]\left | J\right \rangle\\
    &+\prod_{n=1}^{N}i\sin{\theta_n(t)  }\exp\left[-i(\phi_n-\omega_{n,n-1}\tau_{n})\right]\left | J_{\mathrm{max}}\right \rangle,
\end{aligned}
\end{equation}
where the sub-pulse area is defined as  $\theta_n(t)=\mu_{n,n-1}\left|\int_{t_{0}}^{t}E_{n}(t')\exp(-i\omega_{n,n-1}t')dt'\right|$, with the dipole transition matrix element $\mu_{n, n-1}$ and the transition frequency $\omega_{n,n-1}=\omega_n-\omega_{n-1}$ from the $n$th rotational state to the $(n-1)$th rotational state. The values of the transition dipole matrix element between $|J+1\rangle$ and $|J\rangle$ are given by $\mu_{J+1, J}=\mu_0\mathcal{M}_{J+1,J}$, with the permanent dipole moment $\mu_0$ and the transition matrix element $\mathcal{M}_{J+1,J}=(J+1)/\sqrt{(2J+3)(2J+1)}$.\\ \indent 
Based on Eq. (\ref{psi}), we can analyze how the control parameters are mapped onto the wave function of rotational states. As an example, we  consider the $n$th subpulse that has a Gaussian frequency distribution given by $
A_n(\omega)=a_n\exp[-(\omega-\omega_n)^2/2\Delta\omega_n^2]
$, where $a_n$ represents the amplitude, $\omega_n$ the center frequency, and $\Delta\omega_n$ the bandwidth. The time-dependent pulse sequence $E(t)$, capable of generating the coherent superposition of rotational states as described in Eq. (\ref{psi}), can be obtained by
\begin{equation}\label{Et}
   E(t)=\sum_{n=1}^{N}\sqrt{\frac{2}{\pi}}\frac{1}{T_n}\frac{\theta_n(t_{f}) }{\mu_{n,n-1}}e^{-\frac{\left(t-\tau_{n}\right)^2}{2T_n^2}}\cos\left [ {\omega_{n,n-1}(t-\tau_{n})+\phi_n} \right],
\end{equation}
with the subpulse duration $T_n=1/\Delta\omega_n$ and the electric field's amplitude $\mathcal{E}_n=\sqrt{2/\pi}\theta_n(t_f)/(T_n\mu_{n,n-1})$ compared with its definition in Eq. (\ref{pulse}). Through precise modulation of the control parameters, it becomes possible to manipulate an arbitrary high-dimensional superposition of rotational states described by Eq. (\ref{psi}) to  a desired target state defined in Eq. (\ref{target}). This approach can potentially manipulate rotational states, creating a larger state space to store and process information for high-dimensional quantum simulation.\\ \indent
\emph{Application to the Generation of Desired Molecular Orientation.---} As an example, we now analytically design a pulse sequence described by Eq. (\ref{Et}) to  maximize the degree of orientation within a subspace of rotational states with optimal state parameters $\{c_n, \varphi_n\}$ by optimizing control parameters $\{\mathcal{E}_n, \phi_n, \tau_n\}$. By considering the rotational excitations, the degree of orientation at the final time $t_f$ can be quantified by
\begin{equation}\label{<>}
    \left \langle \cos\theta \right \rangle(t_f) = 2\sum_{J=0}^{J_{\mathrm{max}}-1}c_{J+1} c_{J}\mathcal{M}_{J+1,J}\cos{\left(\omega_{J+1,J}t_f-\varphi_{J+1,J}\right)},
\end{equation}
where $\theta$ represents the angle between the field polarization direction and the molecular axis, as depicted in Fig. \ref{fig1}  and $\varphi_{J+1,J}=\varphi_{J+1}-\varphi_{J}$ represents the relative phase difference between $\left| J+1 \right\rangle$ and $\left| J \right\rangle$. The values of $\langle\cos\theta\rangle(t_f)=\pm1$ indicate perfect orientation and perfect anti-orientation, respectively, while an isotropic distribution signifies $\langle\cos\theta\rangle(t_f)=0$.\\ \indent  
 Due to the complexities involved in matrix diagonalization, deriving the analytical eigenvalues and eigenvectors of the operator $\cos\theta$ becomes increasingly challenging for systems that are more complex than simple two- and three-level systems. By utilizing the method of Lagrange multipliers \cite{2004max_Sugny,2020PRA_Wang},  the maximum degree of orientation, i.e. $\lambda\equiv|\left\langle \cos\theta \right \rangle|_{\mathrm{max}}$,  at full revivals for a given value of $J_{\mathrm{max}}$ can be determined by solving the following equation  \cite{SM}
\begin{equation}\label{max}
\lambda^{J_{\mathrm{max}}+1}+\sum_{k=1}^{\left[\frac{J_{\mathrm{max}}+1}{2}\right]}(-1)^k \left\{ \prod_{n=1}^{k}\sum_{\substack{l_1=0;\\l_{n}=l_{n-1}+2}}^{J_{\mathrm{max}}-1-2k+2n}\mathcal{M}_{l_{n}+1,l_{n}}^2\right\}\lambda^{J_{\mathrm{max}}+1-2k}=0,
\end{equation}
where $\left[\frac{J_{\mathrm{max}}+1}{2}\right]$ represents the greatest integer less than $\frac{J_{\mathrm{max}}+1}{2}$. The state amplitudes and phases in the target subspace by Eq. (\ref{target}) should satisfy the following relations
\begin{equation}\label{popu}
c_{J}=\frac{\lambda^{J}+\sum_{k=1}^{[J/2]}(-1)^k \left\{ \prod_{n=1}^{k}\sum_{l_1=0;l_{n}=l_{n-1}+2}^{J-2k+2(n-1)}\mathcal{M}_{l_{n}+1,l_{n}}^2\right\}\lambda^{J-2k}}{\prod_{l=0}^{J-1}\mathcal{M}_{l+1,l}}c_{0},
\end{equation}
and
\begin{equation}\label{Phi}
\varphi_{J+1}-\varphi_{J}=\left(J+1\right)\left(\varphi_{1}-\varphi_{0}\right)+2k\pi, \quad (k=0, \pm 1, \pm 2), 
\end{equation}
with the constraint $\sum_{J=0}^{J_{\mathrm{max}}}c_{J}^2=1$. \\ \indent 
By combining the optimal amplitudes and phase conditions with the analytical wave function in Eq. (\ref{psi}), we can design the pulse sequence in Eq. (\ref{Et})  with optimal control parameters
\begin{equation}\label{amplitude}
\begin{aligned}
\left\{\begin{matrix}  
\theta_1(t_f) = \arccos{c_0}, \quad \\
\theta_n(t_f) = \arccos{\frac{c_{n-1}}{\left | \prod_{k=1}^{n-1}i\sin{\theta_k(t_f)  }\exp\left[-i(\phi_k-\omega_{k,k-1}\tau_{k})\right]\right |}}, \quad (1 < n \leq N),  
\end{matrix}\right.
\end{aligned}
\end{equation}
and
\begin{equation}\label{phase}
\phi_n=\omega_{n,n-1}\tau_{n}- n\left(-\phi_1+\omega_{1,0}\tau_{1}\right)-\frac{(n-1)\pi}{2}+2k\pi, \quad (n > 1),
\end{equation}
for any initial phase $\phi_1$. We can use the optimal pulse area $\theta_n(t_f)$ to calculate the control parameter $\mathcal{E}_n$ of the $n$th subpulse in Eq. (\ref{pulse}). This approach explicitly establishes a mathematical mapping that allows for constructing a desired superposition of multi-rotational states in the target subspace by controlling three parameters $\left\{\mathcal{E}_n, \phi_n, \tau_n \right\}$, which can maximize the degree of orientation at full revivals for a given value of $J_{\mathrm{max}}$.\\ \indent
\begin{figure*}[ht]
\centering
\resizebox{0.9\textwidth}{!}{%
\includegraphics{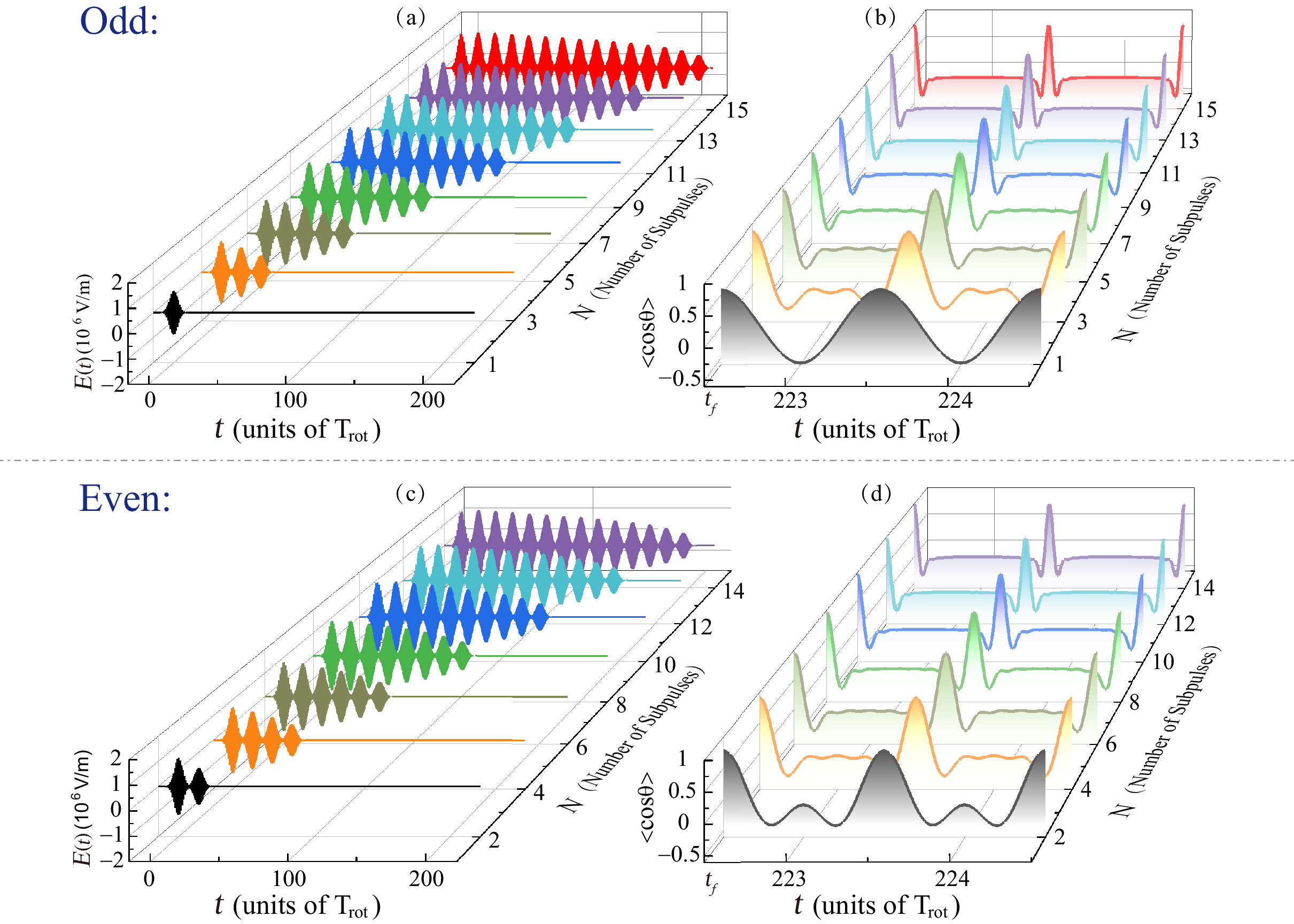} }\caption{The analytically designed pulse sequence (a,c) and  the corresponding degree of orientation $\left \langle \cos\theta \right \rangle$ (b,d) within different subspaces with odd and even numbers of subpulses $N=J_{\mathrm{max}}$. The time is measured in units of the full revival period $T_{\mathrm{rot}}=\pi/B=2.2$ ps of the LiH molecule. All pulse sequences used in our simulations  are turned on at the initial time $t_0=-15T_{\mathrm{rot}}$ and off at the final time $t_f=222.5T_{\mathrm{rot}}$ marked in (b,d).  }
\label{fig2}
\end{figure*}
\emph{Numerical Simulations for Ultracold Polar Molecules.---}  We conduct numerical simulations for ultracold polar molecules initially in the rovibrational ground state of the ground electronic state, which can be obtained under the current experimental conditions \cite{2010Nature_DeMille,2010RMP_Chin,2016PRL_Guo}.  As an example, we consider ultracold LiH molecules with a rotational constant $B=7.513$ cm$^{-1}$ and a permanent dipole moment $\mu_0=5.88$ D. Since our method requires each subpulse with a sufficient long duration $T_n$ for satisfying the resonance conditions, we fix the duration of all subpulses at $T_n=3T_{\mathrm{rot}}$ with $T_{\mathrm{rot}}=\pi/B=2.2$ ps \cite{SM}. By  fixing the center time of $\tau_n=5(n-1)T_n$, we use Eqs. (\ref{amplitude}, \ref{phase}) to determine the control parameters $\left\{\mathcal{E}_n,  \phi_n\right\}$ for a given value of $N=J_{\mathrm{max}}$, which are then substituted into Eq. (\ref{Et}) to obtain the analytically designed pulse sequence $E(t)$. To calculate the time-dependent rotational wave function $|\psi(\theta,t)\rangle$,  we  numerically solve the corresponding time-dependent Schr\"{o}dinger equation governed by the Hamiltonian $\hat{H}(t)=B\hat{J}^2-\mu_0E(t)\cos\theta$ \cite{2010JCP_Shu}. The degree of orientation $\langle\cos\theta\rangle(t)$ is then calculated by $\langle\cos\theta\rangle(t)=\langle\psi(\theta, t)|\cos\theta|\psi(\theta, t)\rangle$, without making use of the approximations employed in our analytical derivations.\\ \indent 
Figure \ref{fig2} shows the time-dependent degree of orientation $\langle\cos\theta\rangle(t)$ induced by the analytically designed pulse sequence, as described in Eq. (\ref{Et}), with $N$ ranging from 1 to 15 in odd numbers for (a,b) and even numbers for (c,d) to avoid the clutter of multiple lines. As depicted in the plot, the analytically designed pulse sequence with a given number of subpulses $N$ in Figs. \ref{fig2}(a) and (c) can achieve the desired maximum degree of orientation,  aligning with the theoretical values derived from Eq. (\ref{max}).  The maximum degree of orientation for one pulse is 0.5774 by including the lowest two rotational states, which has been experimentally demonstrated \cite{2015PRL_Trippel} by using a long nonresonant laser pulse and a weak static electric field. By increasing the number of subpulses in the pulse sequence and modulating their amplitudes and phases, the maximum degree of orientation gradually increases. When we use such a pulse sequence that consists of 15 subpulses to optimize the superposition of the lowest 16 rotational state, the maximum degree of orientation can reach 0.99, approaching perfect orientation. The time-dependent orientation behavior induced by the designed pulses exhibits quantum revivals at regular intervals of $T_{\mathrm{rot}}=\pi/B=2.2$ ps, as shown in Figs. \ref{fig2}(b) and (d). These revivals highlight the periodic nature of rotational dynamics, emphasizing the recurrent patterns in the quantum behavior of the system.\\ \indent 
\begin{figure*}[ht]
\centering
\resizebox{0.9\textwidth}{!}{%
\includegraphics{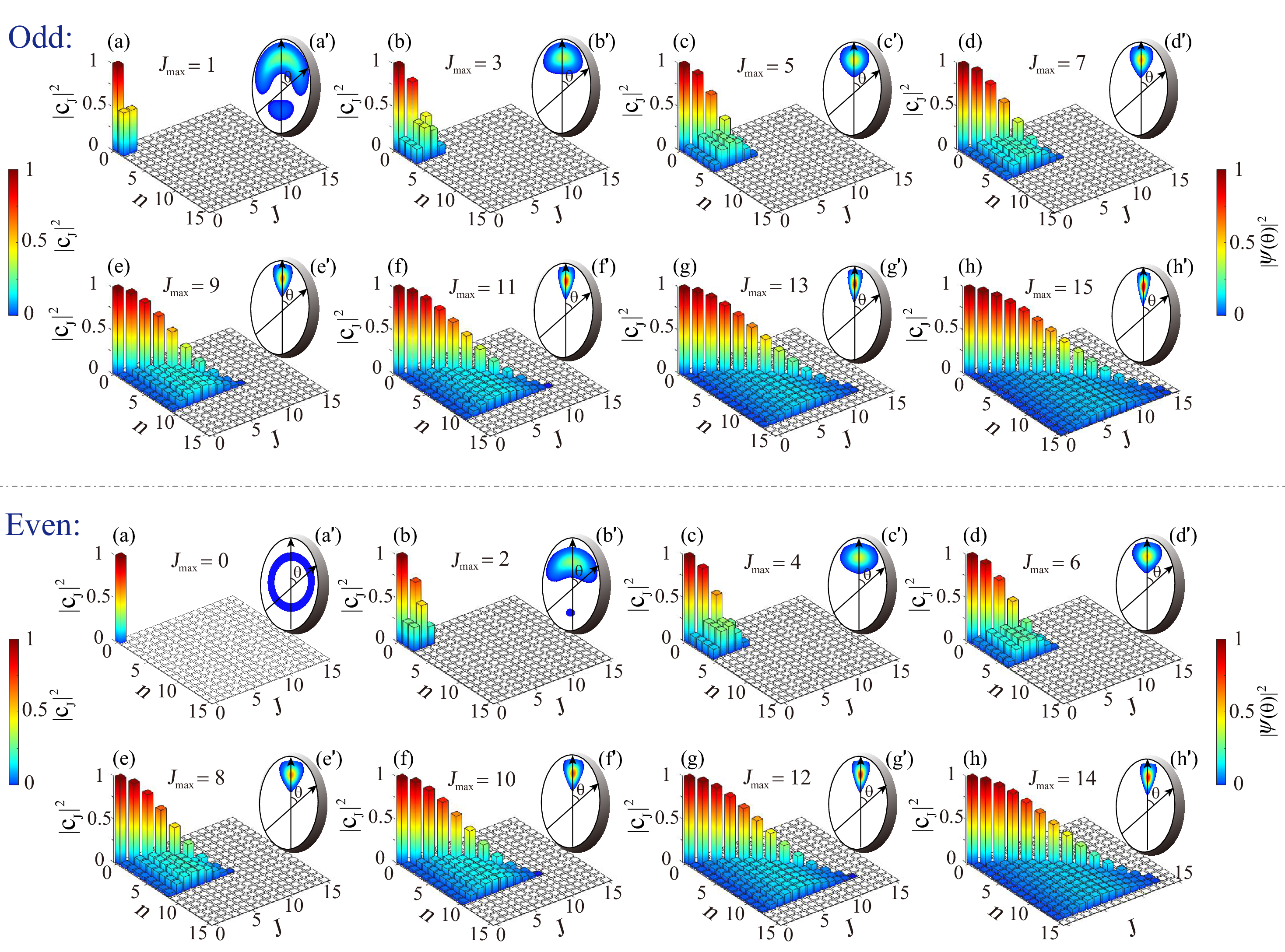} }\caption{The population distributions of $\left | c_J \right |^2$ in (a)-(h) and the corresponding angular distributions of the rovibrational wavepackets in (a')-(h') after the $n$th subpulse excitation for pulse sequences after the $n$th subpulse excitation containing an odd number of subpulses in the upper panels and an even number in the lower panels.}
\label{fig3}
\end{figure*}
Figure~\ref{fig3} illustrates the corresponding rotational population and wavepacket distributions of molecules after interacting with each subpulse under the different pulse sequences used in Fig. \ref{fig2}. Initially, all molecules start in the same rotational state $\left |0\right\rangle$ with $\langle\cos\theta\rangle=0$. The population transfer from the initial state to higher excited states varies with the number of subpulses for a given  pulse sequence. Each subpulse modifies the populations of  adjacent rotational states and adjusts their phases. By applying different pulse sequences to molecules, distinct population distributions of rotational states are achieved, resulting in the desired coherent superposition (wavepacket) of rotational states, which are in excellent agreement with the theoretical values (listed in Ref. \cite{SM}) calculated by using Eq. (\ref{popu}). Since the phases of all rotational states are modulated to the same values as the initial state, i.e., $\varphi_1-\varphi_0=0$ in Eq. (\ref{Phi}), all analytically designed pulse sequences in Figs. \ref{fig2}(a) and (c) drive the molecules to the positive maximum degree of orientation at the same target time, exhibiting revivals to their maximums  at the same delays, as depicted in Figs. \ref{fig2}(b) and (d). Consequently, our simulations clearly demonstrate that these analytically designed pulses can effectively manipulate the rotational motions of molecules, guiding the molecules towards the desired orientation. Our numerical simulations that did not use approximations on the unitary time evolution show excellent agreement with the theoretical predictions,  confirming the validity of the analytical pulses derived from the first-order Magnus approximation. \\ \indent 
\begin{figure*}[ht]
\centering
\resizebox{0.9\textwidth}{!}{%
\includegraphics{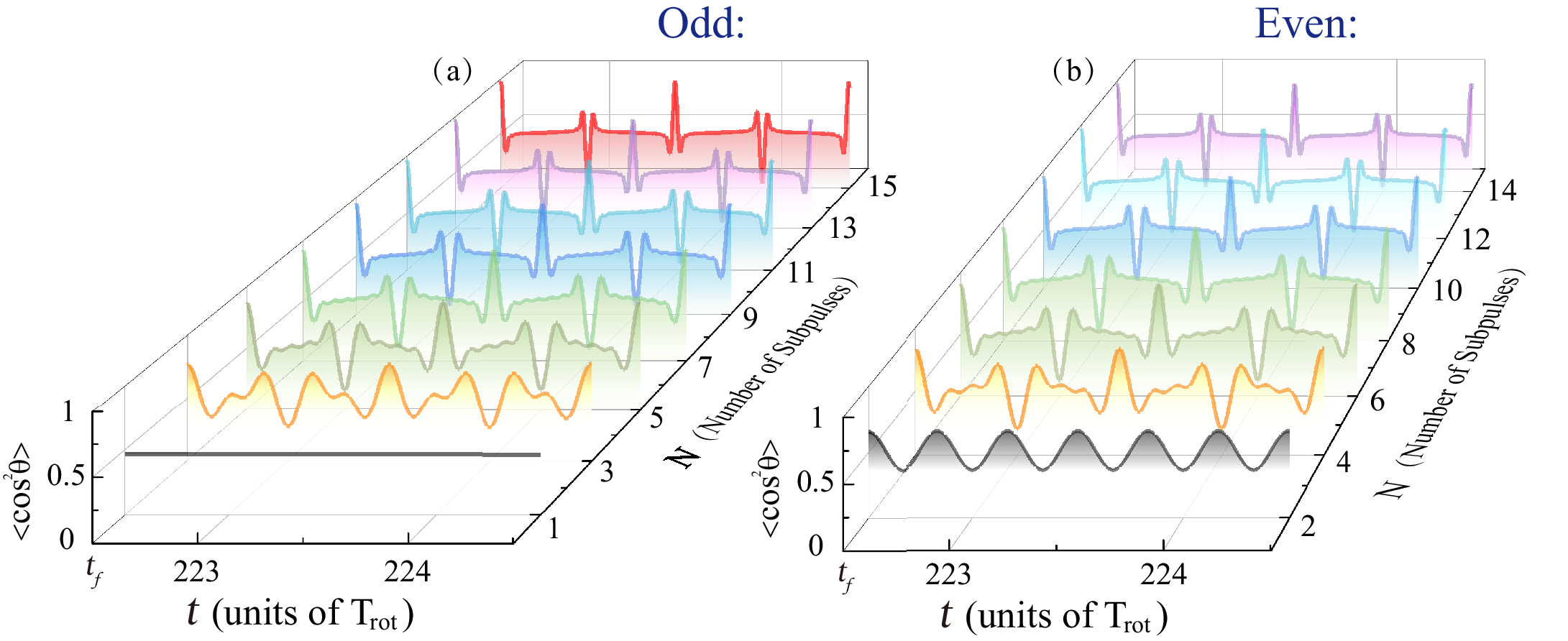} }\caption{The degree of alignment $\left \langle \cos^2\theta \right \rangle$ for different (a) odd and (b) even numbers of $N=J_{\mathrm{max}}$. The time is measured in units of the full revival period $\pi/B$ of the molecule.  The applied pulse sequences are the same as those used in  Fig. \ref{fig2}. }
\label{fig3oe}
\end{figure*} 
We also examine how the analytically designed pulse sequences in Figs. \ref{fig2}(a) and (c) influence the time-dependent degree of alignment $\langle\cos^2\theta\rangle(t)$, as illustrated in Fig. \ref{fig3oe}. In the case of a single pulse, no even rotational states are excited, so we observe that the alignment maintains a constant value of 0.47, below the theoretical maximum of 0.6. As the number of pulses increases, the degree of alignment improves, and periodic revivals occur. Interestingly, a high alignment value over 0.98 can be obtained by exciting the lowest 16 rotational states. It is important to note that the optimized phases between states used in our simulations to achieve the theoretical maximum orientation  are also ideal for attaining the theoretical maximum alignment. To reach the maximum alignment for different values of \( J_{\text{max}} \), we need to  optimize the population distribution of the involved rotational states by only adjusting the strength of each sub-pulse.\\ \indent 
\emph{Discussion.---} Over the past three decades, considerable efforts have been dedicated to generating post-pulse orientations \cite{2012PRA_Tehini,2013PRA_Shu,2014PRL_Kraus,2023PRA_Niels}. These include the development of theoretical models to effectively describe this phenomenon, as well as the design of coherent and optimal control schemes for achieving a high degree of orientation in different molecules \cite{2004PRL_Ortigoso,2012JPCA_Ohtsuki,2013PRA_Rabitz,2014JMO_Koch,2015PRA_Ding,2024PRR_IlyaSh}. The method  being used here requires the involvement of resonant interactions of  tailored terahertz or microwave pulses with the permanent dipole moment of the molecule, in contrast to the nonresonant interactions of  strong laser pulses with the polarizability and hyperpolarizability \cite{2012PRL_Spanner,2014PRL_Kling,2016PRL_Fleischer,2023PCCP_hong}.  Advanced terahertz and microwave pulse-shaping techniques, such as amplitude modulation, frequency modulation, and pulse width modulation, can create desired pulse sequences \cite{2003PR_Goswami,2011OC_Yao,2011PRL_Milner,2014ACP,2021OL_Zeng,2022_Edouard,2023OL_Zhou}. It is also necessary to prepare and measure molecules in specific rovibronic and hyperfine states. Several experimental works  have demonstrated the capability to prepare ultracold molecules in their absolute ground state \cite{2018PRA_Christoph,2022Science_Pan,2024NP_Wang}. Therefore, two necessary conditions required for the realization of this theoretical proposal are technically available. Additionally, the peak intensity of the applied pulse sequence (less than \(2.65 \times 10^5 \, \text{W/cm}^2\) for LiH molecules) is considerably lower than that utilized in nonresonant optical excitation schemes. This indicates that electronic excitation or ionization resulting from non-linear effects could be negligible.
\\ \indent
The maximum degree of orientation achieved depends on the involved  number of rotational states, which in turn determines the required number of subpulses. Increasing subpulses  may lead to molecules susceptible to the negative effects of quantum decoherence on the phases of rotational states. Additionally, including high-lying rotational states may lead to non-negligible coupling of molecular vibrations to the rotational motions \cite{2011PNAS_Yuan}. By  talking into account the centrifugal distortion term into the Hamiltonian, we find that the effect of this term was less than \(10^{-3}\) for \(J_{max}=15\) and is therefore not considered in our simulations. The optimal pulse sequence consisting of 15 subpulses has led to a maximum degree of orientation $|\langle\cos\theta\rangle|_{\rm{max}}>0.99$, which is adequate for practical applications  \cite{2005PRL_Daems}. For the desired target of $J_{max}>15$, we can extend the theoretical method to incorporate the centrifugal distortion term by adjusting the pulse center frequencies while maintaining the other variables  unchanged.
\\ \indent
The present method is universally applicable to different diatomic, linear, and polyatomic  molecules at ultracold temperatures \cite{SM}, whose values of the rotational constants  and the permanent dipole moments determine the amplitude and central frequency of each subpulse. There no added complexities in the optimal pulse sequence for polyatomic nonlinear molecules initially in the absolute ground state ($\nu=0$ and $J=0$).\\ \indent 
The successful application of this analytical method to precise quantum control  of molecular rotation not only showcases its ability to attain maximum molecular orientation values but also underscores its versatility across 15 different scenarios, resulting in the generation of 15 distinct  and  desired  superposition states. These results highlight the potential of applying the proposed analytical approach to precisely construct a desired coherent superposition within a defined subspace of quantum states. In commonly used qubit schemes, one of the main challenges is achieving scalability \cite{2020FP_Wang}. This entails engineering a considerable number of individually controllable qubits within a large Hilbert space that is protected from external perturbations and losses. To address this scalability problem, higher-dimensional quantum systems, known as qudits, have been explored as an alternative to qubits \cite{2005PRL_Neves,2020NJP_Sawant,Ringbauer2022}. Comparing the same size of the Hilbert space, the number of $N$-level qudits needed is notably smaller than the number of qubits, typically reduced by a factor of $\log_2 N$ \cite{2005JMO_Shapiro}. The optimized pulse sequence in Eq. (\ref{Et}) in principle can be potentially applied to the production of qudit gates consisting of multiple quantum states with high-fidelity. \\ \indent
\emph{Conclusion.---} We established an analytical method  called the multi-level pulse-area theorem to effectively control a specific number of quantum states, enabling the construction of desired superpositions of rotational states in molecules. Our results demonstrated how the analytically designed laser pulse sequences can be accurately mapped onto the rotational states within a given subspace.
To illustrate this method, we conducted numerical simulations to precisely control  ultracold polar molecule. The analytically designed pulse sequences can optimize the distribution of rotational populations and the relative phases between adjacent rotational states, leading to the desired molecular orientation. The realization of this theoretical proposal is promising by using current terahertz and microwave pulse-shaping techniques. \\ \indent
The potential application of this analytical framework extends beyond merely achieving maximum molecular orientation. It offers an alternative method for precisely controlling a specific number of rotational states, ultimately enabling the creation of any desired rotational superposition. This capability is vital for utilizing molecular-based qubit, qutrit, and qudit systems in quantum information techniques. This work is essential for preparing high-fidelity coherent superpositions of higher-dimensional quantum states, which has important implications for precision spectroscopy, quantum computing, and quantum information storage and processing in molecules.\\ \indent 
\emph{Acknowledgments.---} This work was supported by the National Natural Science Foundations of China (NSFC) under Grant Nos. 12274470 and 61973317 and the Natural Science Foundation of Hunan Province for Distinguished Young Scholars under Grant No. 2022JJ10070. D. D. is supported in part by the Australian Research Council’s Future Fellowship funding scheme under Project FT220100656. F.N. is supported in part by: Nippon Telegraph and Telephone Corporation (NTT) Research, the Japan Science and Technology Agency (JST)[via the Quantum Leap Flagship Program (Q-LEAP), and the Moonshot R\&D Grant Number JPMJMS2061], the Asian Office of Aerospace Research and Development (AOARD) (via Grant No. FA2386-20-1-4069), and the Office of Naval Research (ONR) Global (via Grant No. N62909-23-1-2074). This work was carried out in part using computing resources at the High Performance Computing Center of Central South University.



%

\end{document}